\documentclass[aps, pra, twocolumn, amssymb, showpacs, superscriptaddress]{revtex4-1}

\usepackage{graphicx}
\usepackage{bm}
\usepackage{float}

\newcommand {\TB}		{T_{B}}
\newcommand {\Bos} 	{Bloch oscillations}

\newcommand {\wf} 		{wave function}

\newcommand {\wpp} 	{wave packet}

\newcommand {\BZ} 		{Brillouin zone}

\newcommand{ \braket } 		[2] { \displaystyle \left<{#1}|{#2}\right> }

\begin{document}

\title{Bloch oscillations in lattice potentials with controlled aperiodicity}

\author{Stefan Walter}
\thanks{present address: Dept. of Physics, Univ. W\"urzburg, Germany}

\author{Dominik Schneble}

\author{Adam C. Durst}
\thanks{present address: Photon Research Associates, Port Jefferson, NY}
\affiliation{Department of Physics and Astronomy, Stony Brook University, Stony Brook, New York 11794-3800, USA}

\date{25 March 2010}

\begin{abstract}
We numerically investigate the damping of \Bos{} in a one-dimensional lattice potential whose translational symmetry is broken in a systematic manner, either by making the potential bichromatic or by introducing scatterers at distinct lattice sites. We find that the damping strongly depends on the ratio of lattice constants in the bichromatic potential and that even a small concentration of scatterers can lead to strong damping. Moreover, collisional interparticle interactions are able to counteract aperiodicity-induced damping of Bloch oscillations. The discussed effects should readily be observable for ultracold atoms in optical lattices.
\end{abstract}


\maketitle

%%%%%%%%%%%%%%%%%%%%%%%%%%%%%%%%%%%%%%%%%%%%%%%%%%%%%%%%%%%%%%%%%%%%%%%%%%%%%%%%%%%%%%%%%%%%%%
\section{Introduction}
\label{sec:introduction}
%%%%%%%%%%%%%%%%%%%%%%%%%%%%%%%%%%%%%%%%%%%%%%%%%%%%%%%%%%%%%%%%%%%%%%%%%%%%%%%%%%%%%%%%%%%%%%

The oscillatory motion of particles in a periodic potential, when subject to an external force, was predicted by Felix Bloch in 1928 \cite{Bloch:1928aa}. Bloch oscillations were first observed in the 1990s both in semiconductor superlattices \cite{Waschke:1993aa} and in systems of laser-cooled atoms in optical lattices~\cite{Wilkinson:1996aa,Ben-Dahan:1996aa}. Since then, they have also been studied in atomic quantum gases~\cite{Morsch:2001aa,Roati:2004aa,Gustavsson:2008aa,Fattori:2008aa} and have found applications in ultracold atomic-physics-based precision measurements ~\cite{Battesti:2004aa,Carusotto:2005aa,Ferrari:2006aa,Clade:2006aa}.

\Bos{} are absent in solids due to fast damping from scattering by defects and phonons. Their observation in long-period semiconductor superlattices relies on oscillation periods that are shorter than the characteristic scattering lifetime, and even in those systems, damping of \Bos{} due to disorder is ubiquitous~\cite{Plessen:1994aa,Diez:1998aa}. In contrast, such damping is absent in optical lattice systems which are inherently defect-free, allowing for the observation of a large number of oscillations~\cite{Roati:2004aa,Gustavsson:2008aa,Fattori:2008aa}. Damping can however be induced by the introduction of disorder into the lattice potential, which can in principle be done with various techniques~\cite{Horak:1998aa,Diener:2001aa,Gavish:2005aa,Ospelkaus:2006aa,Wang:2004aa}.  Also, in the case of quantum gases, a damping of Bloch oscillations arises from mean-field interactions between weakly interacting, Bose-condensed atoms~\cite{Wu:2001aa,Holthaus:2000aa,Witthaut:2005aa, Menotti:2003aa,Modugno:2004aa,Gustavsson:2008aa,Fattori:2008aa}.
In the presence of both interactions and disorder, a reduction of disorder-induced damping due to screening of disorder by the mean field has been predicted~\cite{Schulte:2008aa}. Experiments with ultracold atoms thus not only constitute a versatile testbed for behavior expected in solid-state systems but may also display novel effects.

Recently, the damping of \Bos{} in a Bose-Einstein condensate~\cite{Drenkelforth:2008aa} has been observed for an optical lattice with a superimposed randomly corrugated optical field. A related theoretical investigation of disorder-induced damping of \Bos{} has considered the case of Gaussian spatial noise~\cite{Schulte:2008aa}. In this paper, we numerically investigate the dynamics of an atomic wave packet in a potential with a systematically degraded translational symmetry, considering two scenarios: The first is based on the use of a weak bichromatic potential~\cite{Diener:2001aa} of a variable wavelength ratio, and the second considers scatterers (impurities) pinned at single sites of the potential~\cite{Gavish:2005aa}. We find that the damping strongly depends on the ratio of lattice constants in the bichromatic case and that even a small concentration of scatterers can lead to strong damping. We also include effects of the mean-field interaction and find that the rate at which damping of the \Bos{} occurs is reduced, similar to the case of Gaussian disorder~\cite{Schulte:2008aa}. Both effects should be observable experimentally with existing ultracold-atom technology.

This paper is organized as follows: After a brief discussion of fundamental aspects of tilted lattices in Sec.~\ref{sec:bo}, Section~\ref{sec:aperiodicity} investigates the influence of aperiodicity on the damping dynamics of \Bos{}. Sec.~\ref{sec:interaction} addresses the interplay of aperiodicity and the interaction between atoms. Conclusions are given in Sec.~\ref{sec:conclusion}.

%%%%%%%%%%%%%%%%%%%%%%%%%%%%%%%%%%%%%%%%%%%%%%%%%%%%%%%%%%%%%%%%%%%%%%%%%%%%%%%%%%%%%%%%%%%%%%
\section{Tilted periodic potentials}
\label{sec:bo}
%%%%%%%%%%%%%%%%%%%%%%%%%%%%%%%%%%%%%%%%%%%%%%%%%%%%%%%%%%%%%%%%%%%%%%%%%%%%%%%%%%%%%%%%%%%%%%

The Hamiltonian for the motion of a particle in a one-dimensional periodic potential $V(x) = V(x+a)$ with lattice constant $a$ possesses a complete set of eigenfunctions that obey Bloch's theorem $\varphi(x+a) = e^{ika} \varphi(x)$. The corresponding eigenvalues $E_{n}(k)$ are periodic in momentum space and form energy bands, $E_{n}(k+K) = E_n(k)$  (where $n$ is the band index, $\hbar k$ the quasimomentum, and $K = 2\pi/a$ the width of the first \BZ{}).
Under the influence of an externally applied constant force (``tilt'') $F$, the quasimomentum evolves as
$\hbar k(t) = \hbar k_{0} + F t$. Due to the periodicity of the energy bands, this results in oscillations of the particle's group velocity $v_{g,n}(k) = (1/\hbar) \, dE_n(k)/dk$. These oscillations occur with a period $\TB = 2\pi/\omega_B = h/(a F)$ and a maximum displacement
$2 A_{B,n} = \Delta_n/F$ in coordinate space, where $\Delta_n = |E_n(K/2)-E_n(0)|$ is the width of the $n$-th band.

In the following, we consider particles that are confined to the lowest Bloch band, $n$=1, corresponding to sufficiently deep potentials and small enough tilts such that Zener tunneling~\cite{Zener:1934aa} at the band edges is negligible. Using the split-operator method~\cite{Feit:1982aa}, we perform numerical simulations of the dynamics of Bloch oscillations of a Gaussian \wpp
\begin{equation}\label{eqn:bo6}
	\psi(x,t=0) = \frac{1}{(2 \pi\,\sigma^{2})^{1/4}} \, \exp\left[-\frac{x^{2}}{(2 \sigma)^{2}}\right] \,
\end{equation}
that evolves according to the Hamiltonian
\begin{equation}\label{eqn:bo2}
	H = -\frac{\hbar^{2}\partial_{x}^{2}}{2m} \,  + V_0 \cos(K x) + \tilde{V}(x) + F x
\end{equation}
for motion in a tilted periodic potential $V(x) = V_{0} \cos(K x) + F x$ that is modified by a weak additional potential $\tilde{V}(x)$.

%%%%%%%%%%%%%%%%%%%%%%%%%%%%%%%%%%%%%%%%%%%%%%%%%%%%%%%%%%%%%%%%%%%%%%%%%%%%%%%%%%%%%%%%%%%%%%
\section{Effects of modified periodicity}
\label{sec:aperiodicity}
%%%%%%%%%%%%%%%%%%%%%%%%%%%%%%%%%%%%%%%%%%%%%%%%%%%%%%%%%%%%%%%%%%%%%%%%%%%%%%%%%%%%%%%%%%%%%%

\begin{figure}[t]
	\begin{center}
	\includegraphics[width=1\columnwidth]{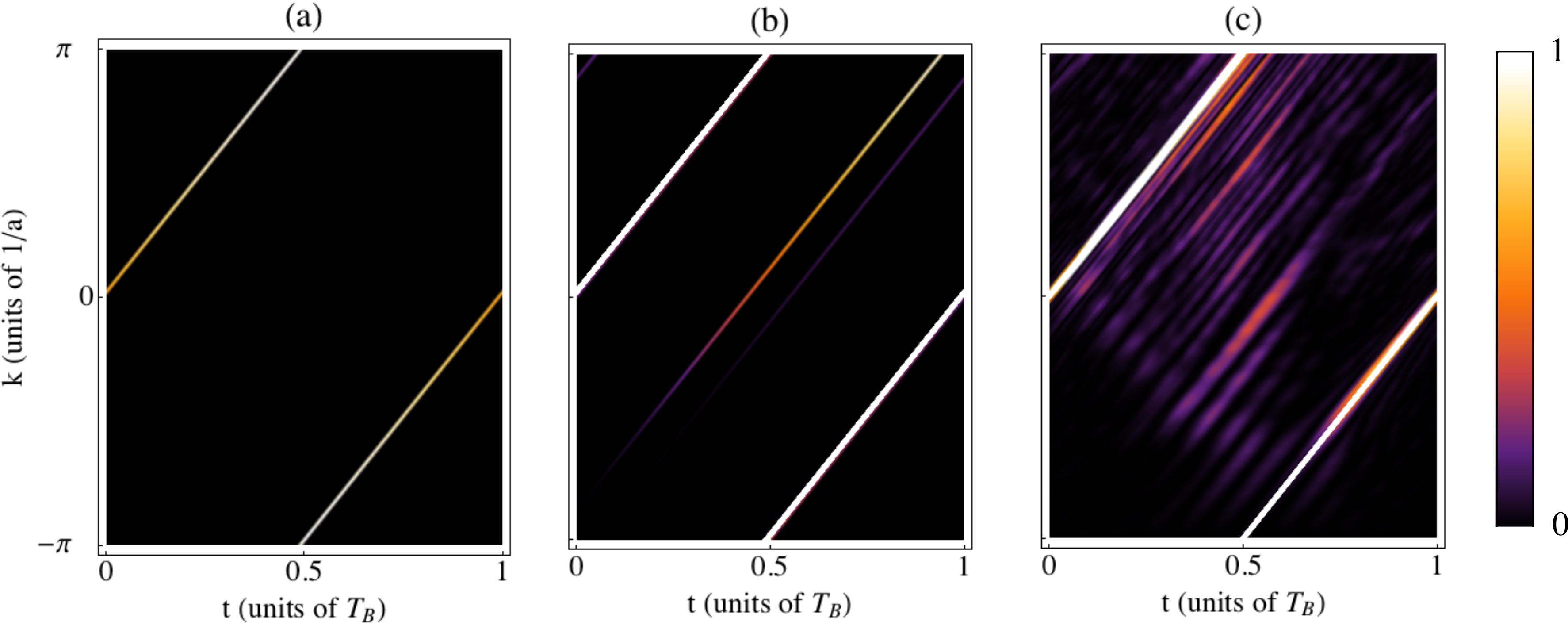}
	\caption{Evolution of the momentum-space density $|\psi(k,t)|^{2}$ (a) in a tilted periodic potential (b) in the presence of an additional periodic potential, and (c) in the presence of two localized scatterers as discussed in the text. Additional momentum components emerge in (b) and (c), broadening the momentum distribution. The parameters in (b) are $\gamma = 0.01$ and $\alpha = 1/\sqrt{5}$; the scatterers in (c) are spaced 11 sites apart. A detailed explanation is given in the text. The density of $|\psi(k,t)|^{2}$ is normalized to $1$; a corresponding color scale is shown on the right.}
	\label{fig:damped_k_dens}
	\end{center}
\end{figure}

This section investigates the influence of the additional potential $\tilde{V}(x)$, which either is a weak additional periodic potential with variable lattice constant, or arises from the local interaction with scatterers pinned at single sites of the tilted periodic potential. In the absence of $\tilde{V}(x)$, the energies of neighboring sites differ by a fixed amount $F a = \hbar\omega_B$, corresponding to a spatially homogeneous phase difference $\Delta\phi(t) = \omega_B t$. In the presence of $\tilde{V}(x)$, $\Delta\phi$ generally becomes position-dependent, leading to global dephasing and thus to a broadening of the wave packet in momentum space, as illustrated in Fig.~\ref{fig:damped_k_dens}.

%%%%%%%%%%%%%%%%%%%%%%%%%%%%%%%%%%%%%%%%%%%%%%%%%%%%%%%%%%%%%%%%%%%%%%%%%%%%%%%%%%%%%%%%%%%%%%
\subsection{Bichromatic potentials}
\label{sec:bichromatic}
%%%%%%%%%%%%%%%%%%%%%%%%%%%%%%%%%%%%%%%%%%%%%%%%%%%%%%%%%%%%%%%%%%%%%%%%%%%%%%%%%%%%%%%%%%%%%%

A tunable bichromatic potential is generated by the addition of
\begin{equation}\label{eqn:bi1}
	\tilde{V}(x) =  \gamma V_0 \cos(\alpha K x),
\end{equation}
with variable relative amplitude $\gamma \ll 1$ and lattice-constant ratio $\alpha$. If $\alpha$ is a rational number $\alpha=p/q$, the total potential has a periodicity $\Lambda = a q$. If furthermore $\Lambda$ exceeds the spatial range probed by the wave packet, the potential can be considered disordered.

\begin{figure}[t]
	\begin{center}
	\includegraphics[width=1\columnwidth]{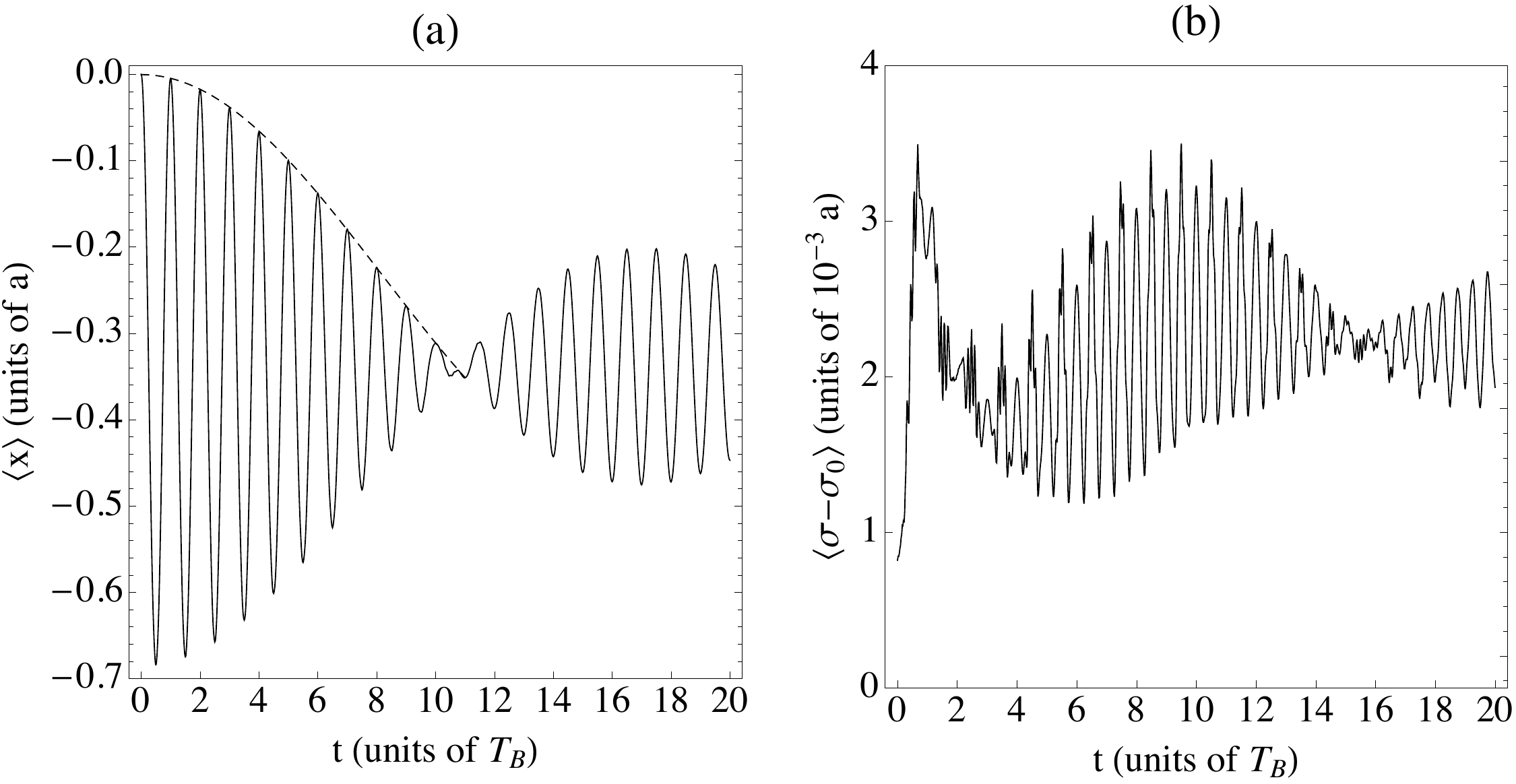}
	\caption{Damped \Bos{} in a bichromatic lattice ($\gamma =0.005$, $\alpha=0.4$). The collapse of the oscillations in coordinate space (a) is accompanied by breathing-mode excitations of the \wpp{} (b), and vice versa (see also text). The dashed line in (a) is the envelope $A \exp(-\eta t^2) + B$.
	\label{fig:dampedBO_0.0005_0.4}}
	\end{center}
\end{figure}

\begin{figure}[t]
	\begin{center}
	\includegraphics[width=1\columnwidth]{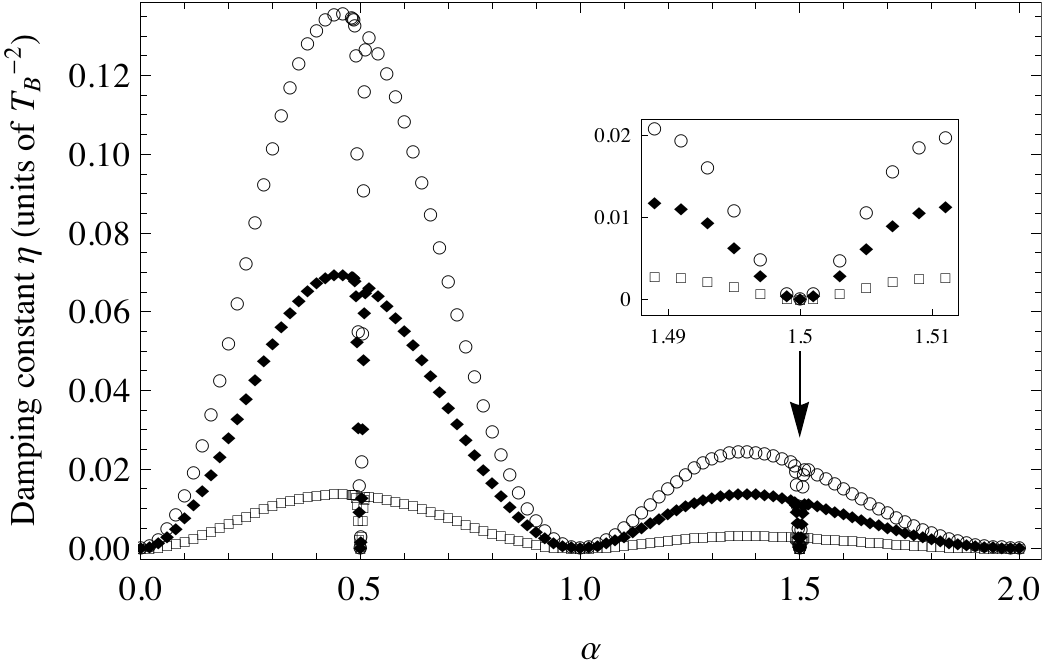}
	\caption{Damping constant $\eta$ as a function of the lattice-constant ratio $\alpha$, for three different depth ratios: $\gamma =0.005$ (squares), $\gamma =0.01$ (diamonds) and $\gamma =0.013$ (circles). The inset shows the behavior of the decay constant in the vicinity of $\alpha = 1.5$.}
	\label{fig:eta_10Er_S_all}
	\end{center}
\end{figure}

The evolution of the \wpp{} (Eq.~\ref{eqn:bo6}) typically exhibits collapses and revivals of center-of-mass oscillations that are coupled to breathing-mode excitations at twice the Bloch frequency as is shown in  Fig.~\ref{fig:dampedBO_0.0005_0.4} (visible after a transient phase immediately following the switch-on of $\tilde{V}$). The presence of a collapse and a revival of the Bloch oscillations is a consequence of the absence of dissipation in our model.
To characterize the decay of Bloch oscillations, the numerical data for their initial decay, for a given $\gamma$ and $\alpha$, are fitted with the function
\begin{equation}\label{eqn:bi2}
	f(t) = A\exp(-\eta t^{2}) \, \cos\left(\omega_B t\right) + B \, .
\end{equation}
Results for the Gaussian-decay constant $\eta$ are shown in Fig.~\ref{fig:eta_10Er_S_all}.

\begin{figure}[t]
	\begin{center}
	\includegraphics[width=1\columnwidth]{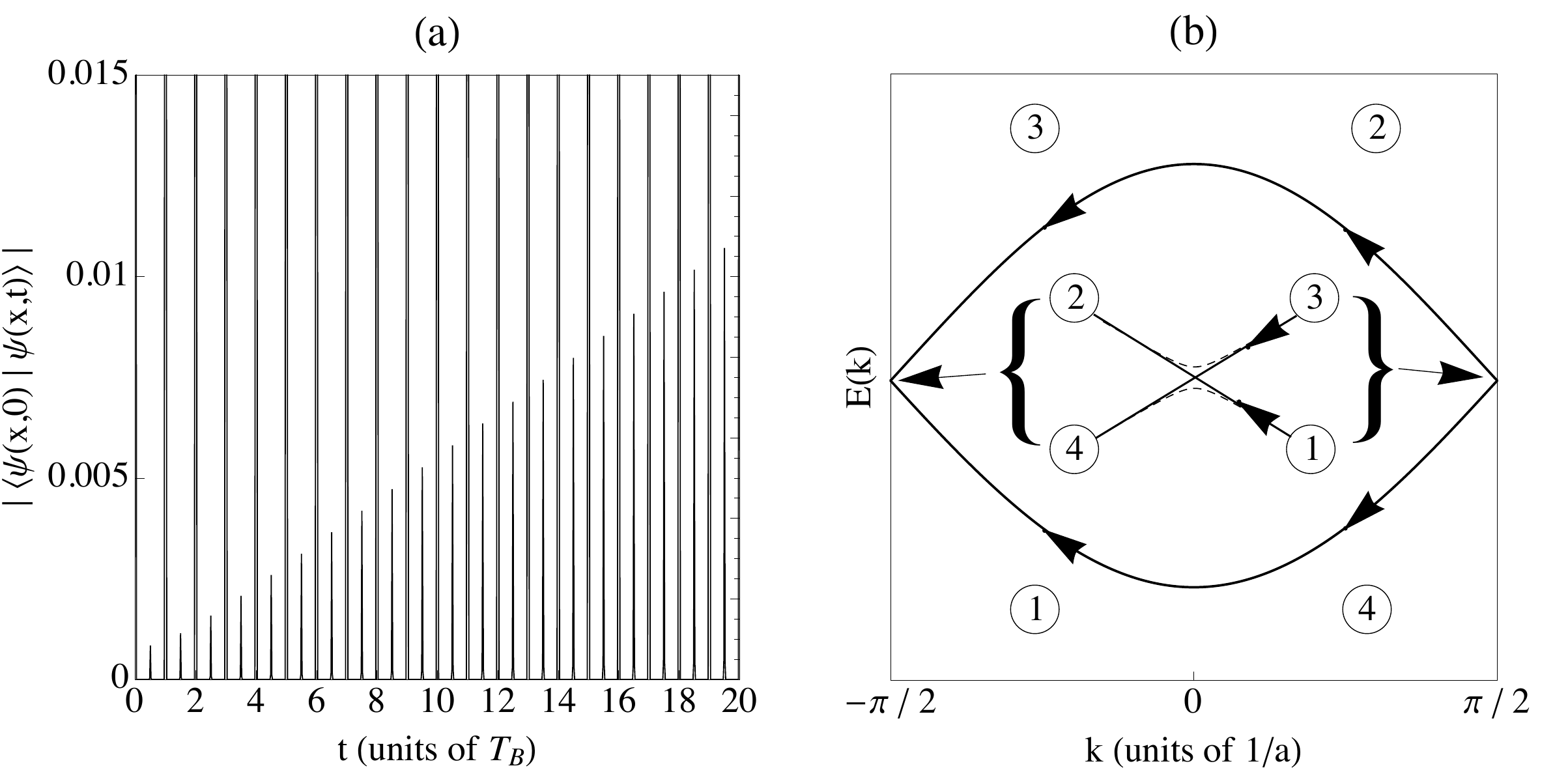}
	\caption{Wave packet in a bichromatic potential with $\alpha=1/2, \gamma = 0.005$. (a) Overlap $|\braket{\psi(x,t=0)}{\psi(x,t)}|$, exhibiting the original periodicity of $T_B$, with small, growing, contributions at odd multiples of $T_B/2$. (b) Band structure of the potential. The inset shows the small gap at the \BZ{} boundaries, with a calculated width $\tilde{\Delta} = 9.5\times10^{-5}~E_{r}$. The arrows with the corresponding numbers indicate the motion of the particle in $k$ space. The wave function completes one Bloch cycle after $T_B$, going from $1 \rightarrow 2 \rightarrow 3 \rightarrow 4$, with most of the wave function tunneling across the gap.}
	\label{fig:alpha05}
	\end{center}
\end{figure}

Clearly an increase in the perturbation amplitude $\gamma$ leads to an overall increase in the damping of the oscillations. However, the dependence of the damping on the ratio $\alpha$ is less trivial. For integer values of $\alpha$, the potential retains its original periodicity and no dephasing of \Bos{} occurs. It is also suppressed for half-integer values of $\alpha$ (i.e. for $q=2$), which correspond to a doubling of the lattice constant, and correspondingly a halving of the \BZ{}, and of the corresponding Bloch period. For this case, the dynamics of the \wpp{} in the modified band structure can easily be visualized. The shape of the band is essentially that of $V(x)$, but it is folded back into the new \BZ{}, thus forming a closed loop, with a small splitting $\tilde{\Delta}\propto\gamma$ at the new zone boundaries (cf. Fig.~\ref{fig:alpha05}). In one Bloch cycle, most of the \wf{} tunnels through the tiny gap from the lower to the upper portion at one boundary and then back to the lower portion at the other boundary (Bloch-Zener tunneling~\cite{Breid:2006aa, Breid:2007aa}). The time needed for one such cycle is $T_B$, that is the Bloch period for the unperturbed potential $V(x)$. This behavior can directly be seen in the time dependence of the overlap of the \wf{} with the initial wave packet $|\braket{\psi(x,0)}{\psi(x,t)}|$, which has a periodicity of $T_B$; contributions at odd multiples of $T_B/2$ only grow very slowly over a large number of cycles.
 
The dynamics of \Bos{} in Figs.~\ref{fig:dampedBO_0.0005_0.4} and \ref{fig:eta_10Er_S_all} are for a wave packet with width $\sigma_0 = 20 a$ in a lattice with depth $V_0 = 10~E_r$ [where $E_r \equiv (\hbar K/2)^2/2m$ is the recoil energy], and a small external force $F = 0.011~V_0/a$.
In coordinate space the amplitude of the \Bos{} is $A_B = 0.34~a$. The additional lattice $\tilde{V}(x)$ is chosen to have a small relative amplitude $\gamma = 0.005$, which is sufficient to cause noticeable damping already after a few Bloch cycles, depending on the lattice constant ratio $\alpha$.

Experimentally, such a bichromatic potential is straightforward to realize in the context of optical lattices, using two laser beams of different wavelength. With present-day tunable-laser technology, a large fraction of the range of $\alpha$ shown in Fig.~\ref{fig:eta_10Er_S_all} can be accessed.

%%%%%%%%%%%%%%%%%%%%%%%%%%%%%%%%%%%%%%%%%%%%%%%%%%%%%%%%%%%%%%%%%%%%%%%%%%%%%%%%%%%%%%%%%%%%%%
\subsection{Scatterers on distinct sites}
\label{sec:scatterers}
%%%%%%%%%%%%%%%%%%%%%%%%%%%%%%%%%%%%%%%%%%%%%%%%%%%%%%%%%%%%%%%%%%%%%%%%%%%%%%%%%%%%%%%%%%%%%%

\begin{figure}[th]
	\begin{center}
	\includegraphics[width=0.85\columnwidth]{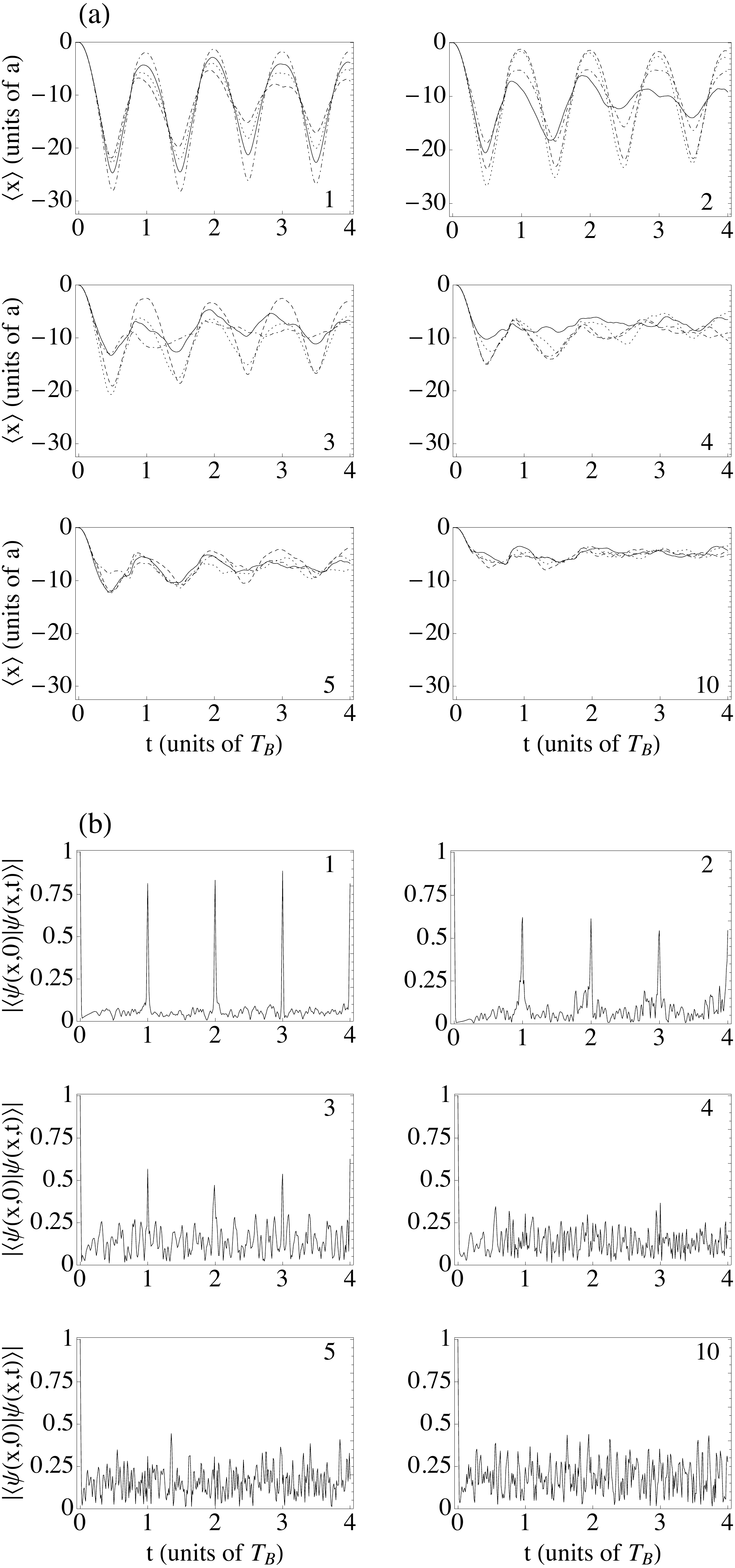}
	\caption{(a) Evolution of the packet position $\left<x(t)\right>$ in the presence of randomly distributed scatterers. Increasing the number of scatterers (as indicated) increases the damping of \Bos. Each curve in a plot represents a different spatial configurations. (b) Overlap with the initial \wf{} for the case of randomly distributed scatterers.  The parameters for (a) and (b) are given in the text.}
	\label{fig:random_grid_1to10}
	\end{center}
\end{figure}

For scatterers pinned at a set of distinct sites $\{n\}$ of the lattice, we model the potential $\tilde{V}$ as a sum of Gaussians,
\begin{equation}\label{eqn:dist1}
	\tilde{V}(x) = \sum_{\{n\}}\tilde{A} \exp\left[-(x-x_n)^2/(2\tilde{\sigma}^2)\right] \, .
\end{equation}
For the simulation, the parameters of the optical lattice and the tilt are chosen such that the size $\sigma$ of the wave packet and its oscillation amplitude $A_B$ in the unperturbed titled potential each cover a large number of lattice sites. A small number of scatterers are then randomly placed within the range of the wave packet's motion. The amplitude $\tilde{A}$ and width $\tilde{\sigma}$ of the scatterers are chosen such that the valleys of the unperturbed potential $V_0 \cos(K x)$ are effectively filled up where the scatterers are located. Their effect on the \Bos{} is shown in Fig.~\ref{fig:random_grid_1to10}. The general trend with an increasing number of scatterers is a more rapid damping of the \Bos{}, with the details of the dynamics depending on their spatial arrangement. Clearly, already a small number of scatterers can lead to a rapid damping of \Bos{}.

The results shown in Fig.~\ref{fig:random_grid_1to10} are for a comparatively shallow lattice ($V_0 = 1.4~E_r$) and a tilting force $F = 0.022~E_r/a$, resulting in a large amplitude $A_B = 16~a$ for undamped \Bos. This condition for the scatterers is well fulfilled by setting $\tilde{\sigma} = \pi a/20 $ and $\tilde{A}=E_r$.

In the context of ultracold atoms, the placement of scatterers pinned at single lattice sites can be achieved, for example by using atoms with two internal states in conjunction with a state-dependent lattice depth~\cite{Gadway:2009aa}, or by using two atomic species in a species-dependent optical lattice~\cite{Catani:2009aa}.

%%%%%%%%%%%%%%%%%%%%%%%%%%%%%%%%%%%%%%%%%%%%%%%%%%%%%%%%%%%%%%%%%%%%%%%%%%%%%%%%%%%%%%%%%%%%%%
\section{Effect of Interactions}
\label{sec:interaction}
%%%%%%%%%%%%%%%%%%%%%%%%%%%%%%%%%%%%%%%%%%%%%%%%%%%%%%%%%%%%%%%%%%%%%%%%%%%%%%%%%%%%%%%%%%%%%%

The discussion so far has been restricted to noninteracting particles. We now consider the case of a Bose-Einstein condensate of $N$ atoms with repulsive interparticle interaction $U(x_i- x_j) = g \delta(x_i-x_j)$. The evolution of the condensate wave function $\psi(x,t) = \sqrt{N}~\phi(x,t)$ in the bichromatic potential is then determined by the mean-field Hamiltonian
\begin{equation}
 \label{eqn:intdis1}
	H = H_0 + \gamma V_0\cos(\alpha K x)+ g N |\phi(x,t)|^{2} \, .
\end{equation}

\begin{figure}[ht]
	\begin{center}
	\includegraphics[width=1\columnwidth]{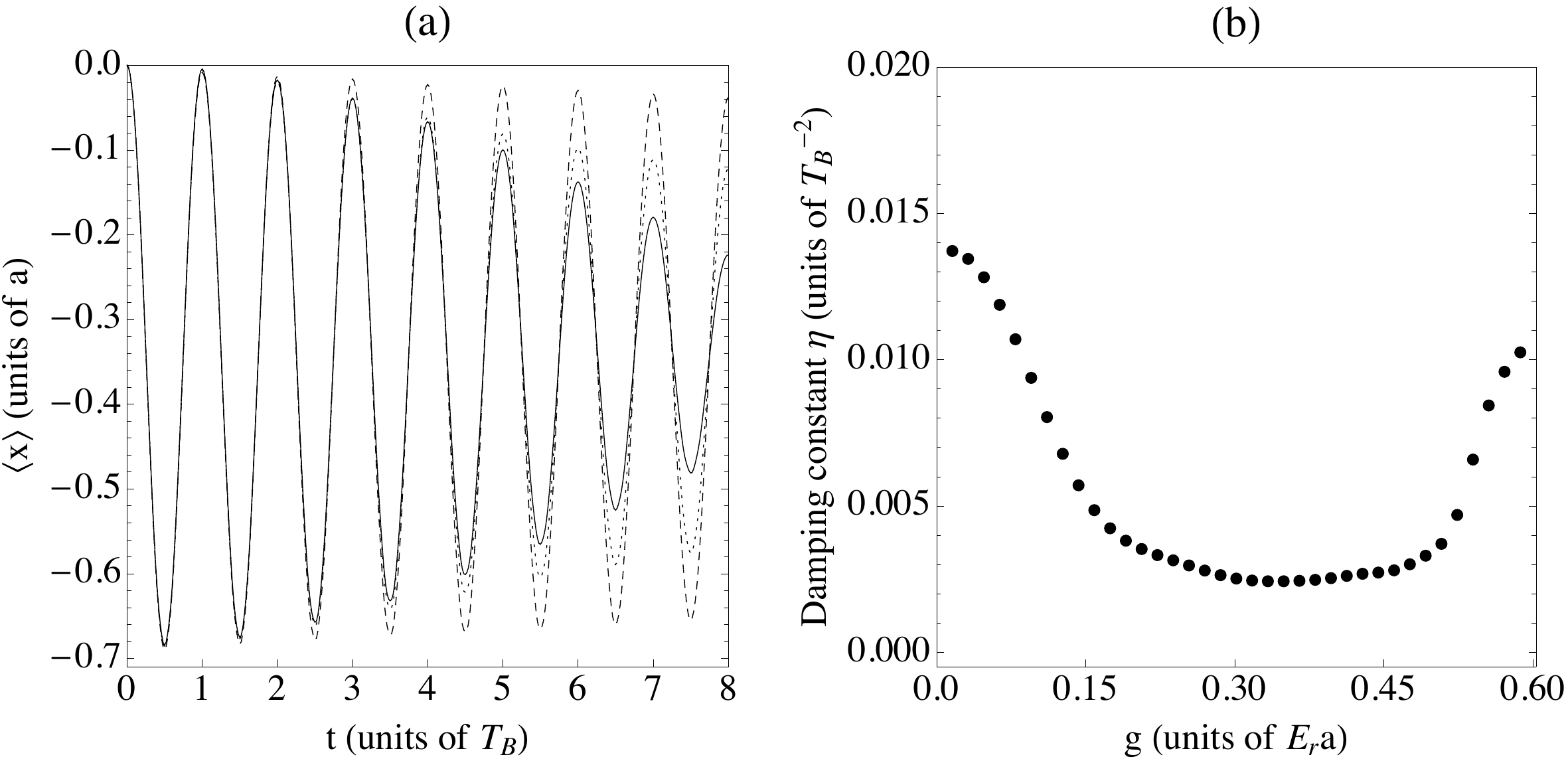}
	\caption{Interplay between aperiodicity and mean-field interaction, showing (a) evolution for $g=0$ (solid), $g=0.1~E_{r}a$ (dotted) and $g=0.4~E_{r}a$ (dashed), and (b) dependence of $\eta$ on the coupling constant $g$, reaching a minimum at $g_c = 0.33~E_r a$}.
	\label{fig:diff_g_paper}
	\end{center}
\end{figure}

The simulation results (Fig.~\ref{fig:diff_g_paper}) show that an increase in the coupling strength $g$ leads to a reduction of the damping constant $\eta$, up to a characteristic value $g_c$, beyond which an increase in $g$ leads to an increase in $\eta$. We interpret the reduction of $\eta$ as a partial screening of the potential corrugations by the mean field~\cite{Schulte:2008aa} as $g$ increases, which is eventually overcompensated by mean-field-induced dephasing~\cite{Holthaus:2000aa,Witthaut:2005aa}.

The parameters for the simulation are the same as those in Sec. 3.1, leading to an associated coupling strength $g_c = 0.33~E_r a$.
This value should be compared with an effective one-dimensional interaction parameter $g = g_{3D}/(2\pi a_{\bot}^{2})$ for a trapped atomic Bose-Einstein condensate, where $g_{3D} = 4\pi\hbar^2 a_s N/m$ (with atomic s-wave scattering length $a_s$ and atomic mass $m$), and $a_{\bot} \sim R$, where $R$ is the Thomas-Fermi radius. Already for a small condensate of $^{87}$Rb atoms with $N=1\times 10^4$ atoms and $R = 5.3~\mu$m in an isotropic 50~Hz trap \cite{Pertot:2009a}, and an optical lattice with $a = 532$~nm, we obtain $g \sim 0.42~E_r a$, which is in the vicinity of $g_c$. Hence, a significant modification of the damping rate in a bichromatic potential due to mean-field effects can be expected; the coupling $g$ depends, for example on the atom number $N$ in the condensate, which is variable. Alternatively, an investigation of the interplay is possible for species in which the mean-field interaction can be tuned via a Feshbach resonance. In this context we mention that \Bos{} with widely controllable mean-field interactions in a (monochromatic) optical lattice have recently been demonstrated with cesium condensates ~\cite{Gustavsson:2008aa,Fattori:2008aa}.

%%%%%%%%%%%%%%%%%%%%%%%%%%%%%%%%%%%%%%%%%%%%%%%%%%%%%%%%%%%%%%%%%%%%%%%%%%%%%%%%%%%%%%%%%%%%%%
\section{Conclusions}
\label{sec:conclusion}
%%%%%%%%%%%%%%%%%%%%%%%%%%%%%%%%%%%%%%%%%%%%%%%%%%%%%%%%%%%%%%%%%%%%%%%%%%%%%%%%%%%%%%%%%%%%%%

We have numerically investigated the damping of \Bos{} resulting from a controlled breakdown of the periodicity of the lattice potential. The effects discussed here, including the effects of the mean-field interaction on disorder-induced dephasing, should be readily observable in experiments with ultracold atoms in optical lattices.

\begin{acknowledgments}
We thank D. Pertot and B. Gadway for a critical reading of the manuscript. This work is supported by NSF Grant Nos DMR-0605919 (S.W. and A.C.D) and PHY-0855643 (D.S.) as well as a DAAD scholarship (S.W.).
\end{acknowledgments}

%\newpage

%%%%%%%%%%%%%%%%%%%%%%%%%%%%%%%%%%%%%%%%%%%%%%%%%%%%%%%%%%%%%%%%%%%%%%%%%%%%%%%%%%%%%%%%%%%%%%

%%%%%%%%%%%%%%%%%%%%%%%%%%%%%%%%%%%%%%%%%%%%%%%%%%%%%%%%%%%%%%%%%%%%%%%%%%%%%%%%%%%%%%%%%%%%%%

\end{document}